# Evaluating Cognitive Assessment Tools: A Comparative Analysis of MMSE, RUDAS, SAGE, ADAS, and MoCA for Early Dementia Detection

Saransh Naole[1], Dhriti Parikh[1], Sakshi Nayak[1], Swarna Priya Ramu[2]

[1] Student, School of Computer Science Engineering and Information Systems

Vellore Institute of Technology naolesaransh@gmail.com

[2] Professor (Grade 1), School of Computer Science Engineering and Information Systems

Vellore Institute of Technology

**Abstract.** Early detection of dementia is very crucial to ensure treatment begins on time, however it is difficult to choose appropriate cognitive assessment tools because each test is designed differently and may not be tailored to the needs of a patient. This review compares five commonly used tests: the Mini-Mental State Examination (MMSE), Rowland Universal Dementia Assessment Scale (RUDAS), Self-Administered Gerocognitive Examination (SAGE), Alzheimer's Disease Assessment Scale (ADAS), and Montreal Cognitive Assessment (MoCA). Each test has different criteria's and vary in their coverage of cognitive domains. MMSE focuses on memory and language but lacks in the evaluation of executive and visuospatial abilities. RUDAS and SAGE focus on memory, language and visual thinking while ADAS mainly targets memory, executive function and language. The MoCA is most complete as it focuses on areas like attention, memory skills, problem solving and visual skills. This review evaluates how accurate and reliable these tools are to help doctors decide the most efficient tool for diagnosis.

## 1   Introduction

Neurodegenerative diseases are marked by the gradual breakdown of neurons in the nervous system [8][9]. Alzheimer's disease (AD) is the most common of these conditions, often leading to memory loss and cognitive decline that significantly affect patients' daily lives [10][13][14][15]. Because early detection is critical for effective treatment and care planning, cognitive assessment tools are central to clinical practice. However, choosing the right tool can be difficult, given differences in design, sensitivity, and applicability.

This review compares five commonly used cognitive assessments for early dementia detection: the Mini-Mental State Examination (MMSE), Rowland Universal Dementia Assessment Scale (RUDAS), Self-Administered Gerocognitive Examination (SAGE), Alzheimer's Disease Assessment Scale (ADAS), and the Montreal Cognitive Assessment (MoCA). The goal is to evaluate each tool's strengths across diagnostic domains, accuracy, and clinical utility, offering guidance for selecting the most suitable option based on specific needs.

## 2      Related Work

Evaluating the effectiveness of cognitive screening tools for dementia and Mild Cognitive Impairment (MCI) has been a focus of significant research. Comparative studies highlight the varying strengths and weaknesses of commonly used instruments. For instance, Dreo et al. [23], using a novel Comparative Specificity and Sensitivity Calculation (CSSC) method, evaluated five tests including MoCA and ADAS. While their findings suggested the brief Phototest offered excellent specificity and efficiency in their cohort, MoCA emerged with the highest sensitivity (95%) but lower specificity (55%), illustrating a common trade-off. Chun et al. [24] conducted a broad scoping review, confirming MoCA and MMSE as frequently studied tools with notable psychometric properties, while also acknowledging the Phototest's specificity and SAGE's utility for self-administration.

The accuracy of MoCA for detecting MCI, a critical precursor stage, has been closely examined. Islam et al. [26] performed a meta-analysis focusing specifically on MoCA's diagnostic accuracy for MCI. Their work suggests that the commonly recommended <26 cutoff yields high sensitivity (93.7%) but poor specificity (58.8%), while a lower cutoff (<23) might provide a more balanced profile (73.5% sensitivity, 91.3% specificity), although they caution about potential bias in the included studies. This highlights the ongoing debate regarding optimal cutoffs for specific clinical purposes.

Detecting even earlier stages, like subtle cognitive decline (SCD), presents further challenges. Pan et al. [25] compared the Memory and Executive Screening (MES) tool against MoCA-Chinese Version (MoCA-CV) and MMSE. They found MES potentially more effective for differentiating SCD from normal cognition, likely due to its greater emphasis on executive and memory skills compared to the other screeners. This suggests that tool selection should consider the specific stage of cognitive impairment being targeted.

A crucial consideration is the applicability of these tools across diverse populations. Zegarra-Valdivia et al. [27] reviewed screenings for illiterate and low-education older adults, noting that while MMSE and MoCA are frequently adapted, inherent educational biases can affect performance. Their review highlighted RUDAS as demonstrating exceptional performance in such groups, underscoring the critical need for culturally and educationally appropriate assessment tools rather than relying solely on adapted versions of existing ones.

In specific clinical contexts, such as geriatric oncology, practical considerations are paramount. Tuch et al. [28] reviewed guidelines and found MMSE and MoCA were the most widely recommended tools, but also noted the value of ultra-brief tools like the Mini-Cog for efficiency. Zhuang et al. [22] also compared various tools for MCI screening, suggesting a step-wise approach—using sensitive tools like p-AD8 or IQCODE for initial screening, followed by more precise tools like MoCA or MES for confirmation—could enhance diagnostic reliability. However, they, along with others [23][24][27], emphasize that test performance varies significantly across studies and populations, necessitating careful validation and interpretation within specific contexts.

## 3    Background and Fundamentals

Neurodegenerative diseases involve the progressive loss of neurons, impairing function often through memory loss or cognitive decline [5]. Alzheimer's disease (AD), the most common neurodegenerative cause of dementia, involves pathological hallmarks like amyloid-beta plaques and tau tangles, leading to progressive cognitive decline, memory loss, and functional impairment [11][5].

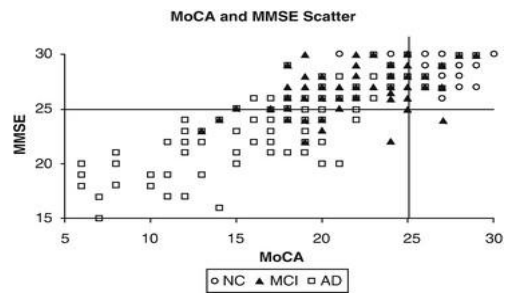

Fig. 1. MMSE and MoCA scores for individuals with MCI and mild AD are plotted in a scatter plot alongside normal controls (NC).

Dementia is defined as cognitive decline (e.g., in memory, language, problem-solving) severe enough to interfere with daily activities. AD is its primary cause, representing a major public health challenge with prevalence expected to rise significantly [5][14].

Mild Cognitive Impairment (MCI) represents a transitional stage between normal aging and dementia, characterized by objective cognitive deficits (e.g., memory loss) without significant impairment in daily functioning. As 10-15% of individuals with MCI may progress to AD annually, early detection using cognitive assessment tools like SAGE or MoCA is critical for monitoring and potential intervention [12][13].

### 3.1 Cognitive Assessment Tools

**MOCA**
A 10-minute screening test for MCI , MoCA assesses executive function, memory, attention, orientation, language, and visuospatial [6][9][10][11][12][13]. It includes subtests such as Alternating Trail Making (drawing numbers/letters without crossing lines for 1 point), Cube/Clock Drawing (drawing a cube or clock for 1 point each), Naming (naming three animals for 1 point each), Memory (recalling five words after two trials with delayed recall scored separately), Attention/Vigilance (repeating digit strings or tapping on "A" in a series for 1 point with ≤1 error), Verbal Fluency (naming≥11 words beginning with "J" in 60 seconds for 1 point), Abstraction (determining
categories for 1 point per pair), and Delayed Recall/Orientation (recalling words and answering date/location questions for points). Scores of ≥26 fall within the typical range. Its multilingual availability, free clinical/educational use, and comprehensiveness are among its advantages [6][9][10]. Formal training, however, might be required for appropriate administration and interpretation [10][12].
Although there is disagreement on subtest equivalence and generalizability, validity is established by its ability to assess many cognitive domains [12]. With 90% sensitivity and 87% specificity, it shows more sensitivity compared to MMSE in detecting MCI and is helpful in differentiating cognitive deficits [9][10][13]. Cultural and educational prejudice is a worry, despite the fact that it has been validated in varied groups [13]. The difference in cognitive performance between people with MCI , mild AD , and normal controls (NC) is shown by Figure 1 [21], which shows the distribution of MMSE and MoCA scores among these groups.

**SAGE**
The Wexner Medical Center at Ohio State University developed the SAGE as a pen-and-paper diagnostic tool to detect cognitive impairments associated with dementia, including AD, in individuals aged 60 and older. [1][2][3][15][16]. SAGE tests a variety of cognitive domains, including communication, thought, and spatial orientation, through a drawing task [1][15][16]. In the comparison of SAGE and MMSE scores across normal, MCI, and dementia subjects (Table I) [21], SAGE showed higher diagnostic accuracy (~95%) compared to MMSE (~90%). In order to minimize practice effects, it offers four variations, enabling longitudinal monitoring and a reference point for comparison. [1][2][3][15].

**Table 1**. Summary of scores

|  | Normal (n=21) | MCI (n=21) | Dementia (n=21) |
|---|---|---|---|
| **SAGE Scores** | *Mean ± SD* | | |
|  | 19.8 ± 2.0 | 16.0 ± 3.2 | 11.4 ± 3.9 |
|  | (Range:22-15) | (Range: 21-9) | (Range: 17-4) |
| **MMSE Scores** | *Mean ± SD* | | |
|  | 28.7 ± 1.1 | 27.7 ± 2.2 | 22.1 ± 3.5 |
|  | (Range:30-26) | (Range:30-23) | (Range: 28-16) |

However, as there is no answer sheet and medical consultation is required for accurate assessment, SAGE must be evaluated by doctors [16]. Scores are influenced by age and education, and people over 80 are advised to raise their score by 1 point [3].

The effects of age and education on SAGE scores are shown in Figures 2 and 3 [3], respectively, emphasizing the necessity of modified scoring thresholds to reduce biases

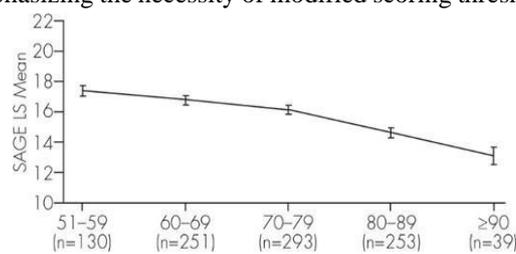

**Fig. 2**. Age Effect on SAGE Score

[3]. Scores of < 17 indicate cognitive impairment, 15–16 indicate moderate mild cognitive impairment, and < 14 indicate dementia [1][3][15][16]. Although a 2–3-point decrease over 12–18 months can indicate early dementia conversion, scores ≤5 have a floor effect, which restricts progression monitoring [3]. SAGE is still a valid early screening instrument despite these constraints, although its findings need to be placed in the overall clinical context.

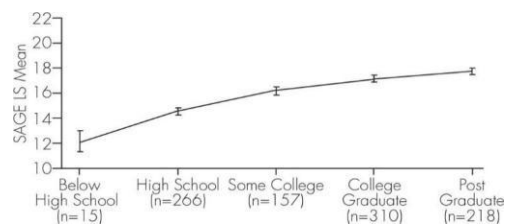

**Fig. 3** Education Effect on Sage Scores

**ADAS-Cog**

A popular tool in clinical trials of AD is the ADAS-Cog, which assesses cognitive domains including language (comprehension, naming), memory (word recall, recognition), orientation, and praxis (e.g., constructional, ideational) [17]. The exam typically takes 30-35 minutes, with scores ranging from 0 to 70 (lower scores indicating better function, although sometimes reversed). Compared to the MMSE, it is considered more sensitive to cognitive deterioration over time in AD, less dependent on education or language, and is thus crucial for evaluating treatment effectiveness in clinical trials [18][20][17][19][15]. Despite strong inter-rater and test-retest reliability, variability in administration and standardization across studies has been questioned [18]. Its length and complexity often require specialist training [18], and a notable limitation is its weak assessment of executive functions or planning [17]. While adaptations exist, the original ADAS-Cog remains a standard but requires careful interpretation, recognizing that clinical stability might not always align with score changes [18].

**RUDAS**

The RUDAS was created in Australia (2004) to facilitate dementia diagnosis in culturally and linguistically diverse (CALD) groups, where standard tests may be biased [14]. Given the rising global burden of dementia, particularly in low- and middle-income countries, culturally fair screening tools are essential. The RUDAS is a brief cognitive exam designed to minimize language and cultural influences. It evaluates several domains including memory, praxis, visuospatial construction, judgment/abstraction, and language aspects like body part identification. Scores range up to 30, with higher scores indicating better cognition. It demonstrates excellent diagnostic utility for dementia and MCI and high reliability (specific metrics in Table 3) across various translations [14]. Although largely independent of language or gender, educational level can still have some impact. The RUDAS serves as an effective screener to identify individuals needing further assessment, rather than a standalone diagnostic tool, and should be administered considerately [14].

**MMSE**

The Mini-Mental State Examination (MMSE), developed by Folstein et al., remains one of the most widely used brief screening tools for cognitive abilities, particularly in older adults [7][8]. Typically taking 10-15 minutes, it assesses domains including orientation (time, place), registration (immediate memory), attention and calculation, recall (delayed memory), language, and basic visuospatial skills (copying) [7]. Scores range from 0 to 30, with higher scores indicating better cognitive function; a common, though debated, cutoff for impairment is <24 or <25 [8]. While rapid and familiar, the MMSE's sensitivity for detecting mild cognitive impairment (MCI) or early dementia is limited, especially compared to tools like the MoCA [8][9]. Performance can be significantly influenced by education level, cultural background, and primary language, requiring careful interpretation [7]. Furthermore, it may exhibit "floor effects" in severe dementia (scores near zero limiting detection of further decline) and notably lacks robust assessment of executive functions [8]. Consequently, while useful as a quick screen, the MMSE is not a diagnostic tool and often requires supplementation with more comprehensive testing, like the MoCA, for suspected mild impairments [8][9].

# 4 Methodology

MMSE, RUDAS, SAGE, ADAS, and MoCA are just a few of the cognitive assessment techniques that are thoroughly covered in this review article. We used a rigorous and methodical approach that comprised crucial stages for locating relevant studies, research articles, and meta-analysis for data collecting and com- parison in order to assess the limitations and drawbacks of the Moca test in comparison to other cognitive testing instruments. [12] When selecting the studies, data, and re- search articles included in comparison analysis, we used certain inclusion and exclusion
criteria. We conducted a thorough search of the literature using the resources that our mentor recommended, including PubMed and Google Scholar. [10] [12]

# 5 Discussion

The MoCA is a frequently used 10-minute test for executive function, language, memory, and attention [4][6][12], valuable for identifying early cognitive impairment. However, its utility must be balanced against its limitations. As highlighted by the comparative analysis (Table II), cognitive assessment tools vary significantly in domain coverage, administration time, and application. While MoCA offers high sensitivity for MCI, factors like training requirements, educational bias, and potential ceiling effects can constrain its use [6][12]. Alternatives like SAGE offer self-administration [1], RUDAS provides a culturally fairer option [12], and ADAS-Cog is tailored for monitoring AD changes in trials [14][15], though the widely known MMSE is less sensitive to mild impairment [12]. Selecting the appropriate tool requires considering the specific patient population and clinical context.

## 5.1 Training and Certification

Since September 2019, practitioners have needed mandatory certification, which includes training and a fee of about $125. While it helps standardize administration, it may also pose a financial or accessibility hurdle for some.

## 5.2 Education and Cultural Bias

Because the MoCA emphasizes executive function and visuospatial skills, individuals from different cultural or educational backgrounds may receive lower scores that don't necessarily reflect true cognitive impairment [6]. Research has noted these discrepancies, underscoring the importance of culturally responsive assessment strategies [6].

## 5.3 Ceiling Effect

In people with high baseline cognition, the MoCA may not be sensitive enough a phenomenon known as the "ceiling effect" [4][12][13].

## 5.4 Limited Specificity

While the MoCA covers general cognitive functions, it isn't tailored to detect issues linked to specific disorders like frontal lobe dysfunction or Parkinson's disease-related impairments [4]. Because of its broad approach, additional neuropsychological tests are often needed to pinpoint specific deficits or track more subtle changes over time [4].

## 5.5 Alternatives to MoCA

Other tools fill in some of the gaps the MoCA might leave. RUDAS is great for culturally diverse populations, since it was built to minimize cultural bias [12]. SAGE is self-administered, so people can take it on their own—even remotely [1]. The MMSE is fast and popular, but not as good at picking up early cognitive decline [12]. And ADAS-Cog, developed for Alzheimer's trials, is especially useful for tracking changes over time [14][15].

## 5.6 Integration of longitudinal data in cognitive assessment:

To effectively track cognitive changes over time, consider incorporating longitudinal data through methods such as:
- **Serial Testing:** Administering the same tool (e.g., MoCA, MMSE, SAGE) at regular intervals.
- **Standardized Change Scores:** Calculating scores that reflect change, accounting for practice and age effects.
- **Composite Scores:** Combining results across multiple cognitive domains for a holistic view.
- **Automated Adaptive Testing:** Using adaptive tests to enhance sensitivity and minimize floor/ceiling effects.

- **Biomarker Integration:** Combining cognitive results with neuroimaging or other biomarkers.
- **Patient-Reported Outcomes:** Including patient-reported functional measures for real-world relevance.
  These approaches enhance the monitoring of cognitive decline and aid early detection of neurodegenerative diseases.

### 5.7 Recommendations for Improving MoCA

To mitigate MoCA's educational sensitivity and cultural biases, consider:
- Developing culturally adapted versions with local norms.
- Using adaptive testing to adjust difficulty based on education level.
- Applying separate cut-offs for memory vs. non-memory tasks, as memory recall appears less education-dependent.
- Utilizing composite scores combining MoCA with other tests.
- Validating cutoffs for each adapted version through specific studies.
- Providing targeted administrator training focused on cultural contexts.
- Employing computerized versions that can potentially adjust for culture and education.

## 6 Conclusion

Despite being translated into around 60 languages, the MoCA has several important drawbacks, such as its sensitivity to schooling, its failure to account for respondents' educational backgrounds, and potential linguistic and cultural biases. People with very high baseline capacities may not be able to identify cognitive loss because to a ceiling effect, and MoCA only measures a single point in time rather than cognitive change across time. In order to avoid misinterpreting results and to use tests with higher sensitivity and specificity (e.g., MMSE, SAGE, RUDAS), physicians must be aware of these limitations in order to make correct clinical and research applications.

By adding supplementary tests, minimizing biases within multi-ethnic populations, and accounting for educational sensitivity, modifications can improve the usefulness of MoCA. Its usefulness in evaluating cognitive change would be increased by adding clauses to track longitudinal performance. MoCA will become a more equal and accessible tool for measuring cognitive function as a result of these modifications.

**Table 2.** Comprehensive comparative analysis of MMSE, RUDAS, SAGE ADAS, and MOCA

| Category | SAGE | RUDAS | ADAS | MMSE | MOCA |
|----------|------|-------|------|------|------|

| Cognitive Domains | Memory, Language, Executive function, Orientation | Memory, Language, Praxis, Visuospatial skills | Memory, Language, Praxis | Orientation, Memory, Attention, Language | Attention, Memory, Language, Executive function |
|---|---|---|---|---|---|
| Admin Time | ~15 min | ~20 min | Up to 60–90 min | ~10–15 min | ~10–15 min |
| Age Group | ≥50 | Older adults+ | Older adults | Older adults | All adults |
| Brain Area Targeted | Cortical/subcortical regions | Various brain regions | Memory/cognition regions | Cortical functions | Broad cognitive regions |
| Self-Admin Capability | Yes | Clinician-administered | Professionals only | Clinician-administered | Clinician-administered |
| Clinical Outcome | Further evaluation needed | Assess cognitive impairment | Elevated scores indicate decline | Indicates impairment | Indicates impairment/MCI |
| AD Detection | Early cognitive decline | Early Alzheimer's detection | Sensitive to Alzheimer's changes | Moderate-severe Alzheimer's | Early Alzheimer's/MCI |
| Certification Process | No certification; self-administered, clinician evaluation needed. | No certification; training recommended for multicultural settings. | No certification; usually clinician-administered, training may be needed. | No certification; training recommended to reduce errors. | Certification required since 2019 ($125); includes standardized training. |

**Table 3.** Evaluation of Cognitive Assessment Tools Based on Key Metrics

| Tool | Sensitivity | Specificity | p-value | Other Metrics |
|---|---|---|---|---|
| SAGE | 95% | 95% (detecting cognitive impairment) | $p < 0.0001$ | Correlation with neuropsychological battery ($r = 0.8$) |
| MMSE | 81% | 89% | $p < 0.001$ | Area under ROC curve: 0.92 |
| MoCA | 90% | 87% | $p < 0.001$ | Area under ROC curve: 0.97, Cohen's $d = 2.15$ |
| RUDAS | 89% | 93% | $p < 0.001$ | Cronbach's alpha: 0.74 |
| ADAS-Cog | 92.20% | 98% (for AD, cut-off >12); 91% (for MCI) | $p < 0.05$ | Test-retest reliability: 0.91, Cohen's $d = 0.3$ to 0.5 |